# Ferroelectric Al$_{1-x}$B$_x$N sputtered thin films on n-type Si bottom electrodes

*Ian Mercer[1], *Chloe Skidmore[1], Sebastian Calderon[2], Elizabeth Dickey[2], Jon-Paul Maria[1]

*denotes equal contribution

[1]Department of Materials Science and Engineering, The Pennsylvania State University, University Park, PA, USA

[2]Department of Materials Science and Engineering, Carnegie Mellon University, Pittsburgh, PA, USA

## Abstract

Ferroelectric Al$_{1-x}$B$_x$N thin films are grown on highly doped and plasma treated (100) *n*-type Si. We demonstrate ferroelectricity for x = <0.01, 0.02, 0.06, 0.08, 0.13, and 0.17 where the *n*-type Si is both the substrate and bottom electrode. Polarization hysteresis reveals remanent polarization values between 130-140 μC/cm$^2$ and coercive field values as low as 4 MV/cm at 1 Hz with low leakage. The highest resistivity and most saturating hysteresis occurs with B contents between x = 0.06 and 0.13. We also demonstrate the impact of substrate plasma treatment time on Al$_{1-x}$B$_x$N crystallinity and switching. Cross-sectional transmission electron microscopy and electron energy loss spectra reveal an amorphous 3.5 nm SiN$_x$ layer at the Al$_{1-x}$B$_x$N interface post-plasma treatment and deposition. The first ~5 nm of Al$_{1-x}$B$_x$N is crystallographically defective. Using the *n*-type Si substrate we demonstrate Al$_{1-x}$B$_x$N thickness scaling to 25 nm via low frequency hysteresis and CV. Serving as the bottom electrode and substrate, the *n*-type Si enables a streamlined growth process for Al$_{1-x}$B$_x$N for a wide range of Al$_{1-x}$B$_x$N compositions and layer thicknesses.

## Introduction

The wurtzite ferroelectric materials family over recent years has encountered significant excitement toward integrated robust ferroelectric memory.[1] First reported by Fitchner et al. in switchable Sc-substituted AlN, followed by Hayden et al. in B-substituted AlN and Ferri et al. in ZnO and MgO solid solutions, a rich landscape remains for discovering "ferroelectrics everywhere".[2–4] AlN based systems can be prepared at low growth temperatures (< 400 °C) that are compatible with back-end processing, using only established CMOS elements (e.g. Al, B, N). These films exhibit remanent polarizations above 100 $\mu C/cm^2$, and can be integrated with mainstream semiconductors and substrates.[5] Furthermore, despite their relative youth these wurtzite ferroelectrics can be scaled below 10 nm, which is a concern in some other materials intended for modern device integration.[6] Typically, growing $Al_{1-x}B_xN$ on sapphire and Si substrates requires bottom electrode preparation that optimizes roughness, minimizes deposition strain, and promotes $c$-axes wurtzite texture. The latter is particularly important given the [0001] $Al_{1-x}B_xN$ polar axis orientation and its uniaxial nature.[5,7,8]

Much of the AlN electrode working knowledge is based on resonating devices that span at least 15 different metal electrodes, many of which significantly affect device performance.[9] For $Al_{1-x}B_xN$ on Si with tungsten (W) bottom electrodes, rocking curve full-width-half-maximum (FWHM) values of ~1.4° and ~ 1.6° are typical for the (011) W and (002) $Al_{1-x}B_xN$ respectively.[5,10] To optimize W crystallinity and texture the Si is first plasma treated to create a nitrogen rich $SiO_xN_y$ surface layer, followed by sputter depositing a thin AlN seed layer before W sputter deposition. Ultimately, it is thought by maximizing bottom electrode quality the electrical quality of the $Al_{1-x}B_xN$ is enhanced.[5] Interfacial defect densities, which appear to erode insulation resistance especially at lower thicknesses, however, remain high in the $Al_{1-x}B_xN$.[11] While high crystalline fidelity is commonly associated with desirable electrical properties, the correlations in ferroelectric wurtzites are not completely clear. For example, the thinnest functional $Al_{1-x}Sc_xN$ capacitors to date are non-epitaxial films on Pt bottom electrodes.[12] The chemical compatibility connecting AlN and Si invites one to consider heavily doped Si as substrate and bottom electrode, a stack that could, in principle, streamline $Al_{1-x}B_xN$ capacitor integration.[13,14]

This manuscript explores $Al_{1-x}B_xN$ (x = <0.01, 0.02, 0.06, 0.08, 0.13, and 0.17) prepared directly on $n$-type doped Si substrates with specific attention to in-situ substrate plasma treatments that promote and optimize crystallinity, texture, and uniformity. In general, we find that $Al_{1-x}B_xN$ thin films can be deposited with nearly the same crystallinity on $n$-type Si as on W provided the correct plasma pre-treatment, and that ferroelectric properties are nearly identical.

## Material Preparation

Sputtered $Al_{1-x}B_xN$ thin films are grown on 4 in. (100) highly As doped (0.001-0.005 Ohm-cm) $n$-type Si substrates (Nova Electronic Materials and University Wafer). The films are sputtered using 2" Kurt J. Lesker MagKeeper cathodes connected to a 2" Al (99.9995% Kurt J. Lesker) and 2" BN (99.5% Plasmaterials) targets. Al and BN targets are both powered by separate Kurt J. Lesker RF R301 generators with auto-tuning match networks. The magnetrons are positioned in a confocal configuration with a 22.5° angle with respect to the substrate normal and a 10 cm target-to-substrate distance for each gun. The substrate heater (AJA International) is modified to allow for deposition on 4" wafers with direct radiative heating of the wafer backside using two 1000 W halogen lamps. Chamber base pressures for depositions are all below 8 x $10^{-8}$ Torr. Before the depositions, substrates are first loaded into the vacuum chamber and heated to 600°C for 30 minutes then cooled down to 350°C for another 30 minutes. Each plasma treatment is performed at 20 mTorr in 20 sccm Ar and 20 sccm $N_2$ with a 50 W substrate bias applied by a Kurt J. Lesker RF R301 Power Supply. Al and BN targets are sequentially pre-sputtered at 10 mTorr with 40 sccm Ar for 5 minutes before each growth. The Al target power is maintained at 300 W for all depositions and the BN target power is set to 20 W, 50 W, 100 W, 150 W, 200 W, and 250 W corresponding respectively to $Al_{1-x}B_xN$ where x = <0.01, 0.02, 0.06, 0.08, 0.13, and 0.17. Note the B content for x = <0.01 is under the detection limits of X-ray photoelectron spectroscopy measurements. Growths are optimized at 350°C and 1.9 mTorr flowing 24 sccm Ar and 16 sccm $N_2$. X-ray reflectometry (XRR) film thickness measurements show that deposition rate increases linearly with respect to the BN cathode power. Deposition times are modified accordingly to achieve the desired thickness values. To complete capacitor stacks, W top electrodes ~50 nm thick and 100 μm diameter are deposited on the $Al_{1-x}B_xN$ films at room temperature using a shadow mask. Note that the films are exposed to atmosphere and moved to another chamber for top electrode deposition. Precise electrode sizes are imaged by an optical microscope and the areas are measured by ImageJ.

A PHI VersaProbe II X-Ray Photoelectron Spectrometer (XPS) is used to quantify the varying composition of each sample in the series. Before XPS, a 2 minute Ar sputter is performed 2keV (2 mm x 2 mm spot size) to reduce contributions from surface oxidation and hydrocarbons. Multiple pass high resolution scans of the Al 2p and B 1s were used to determine B composition. A Panalytical Empyrean x-ray diffractometer (XRD) with a Cu source and five-axis cradle is used to quantify crystallinity and mosaic spread. Every scan uses a Bragg–Brentano[HD] incident optic and a PIXcel[3D] detector with programmable anti-scatter slits. An Asylum MFP-3D Atomic Force Microscope (AFM) in a repulsive tapping mode is used to image surface topography and roughness. An aixACCT TF 3000 Analyzer and an aixACCT 150 V High Voltage Amplifier are used for all polarization hysteresis measurements. An in-situ Film Sense Multiwavelength uv-vis ellipsometer is used to measure refractive indexes and also confirm film thickness with X-ray Reflectivity. Transmission electron microscopy (TEM) images are acquired in a Tecnai F20 operated at 200kV. Electron Energy Loss spectroscopy (EELS) is acquired in a Gatan Tridiem spectrometer with an energy resolution of approximately 1 eV.

**Results and Discussion**

As a first experiment, 200 nm $Al_{1-x}B_xN$ films are grown on *n*-type Si with six different B concentrations: x = <0.01, 0.02, 0.06, 0.08, 0.13, and 0.17, the *n*-type Si wafers are untreated before loading into the deposition chamber. X-ray diffraction and AFM analysis are shown in Fig. 1. Fig. 1(a) shows stacked θ-2θ scans for the entire series zoomed in around the (002) wurtzite reflection. B reduces the *c*-axis lattice parameter consistent to previous work explored by Hayden et al., with a modest peak broadening effect. Fig. 1(b) presents FWHM rocking curve values for the same B series, showing that the out-of-plane mosaic spread increases monotonically with B content, as reported previously.[11] Fig. 1(c) shows AFM topography scans of an as-received *n*-type Si surface and a 200 nm thick $Al_{0.92}B_{0.08}N$ surface; RMS roughness values are 161 pm and 380 pm respectively. The latter is representative of all other $Al_{1-x}B_xN$ films.

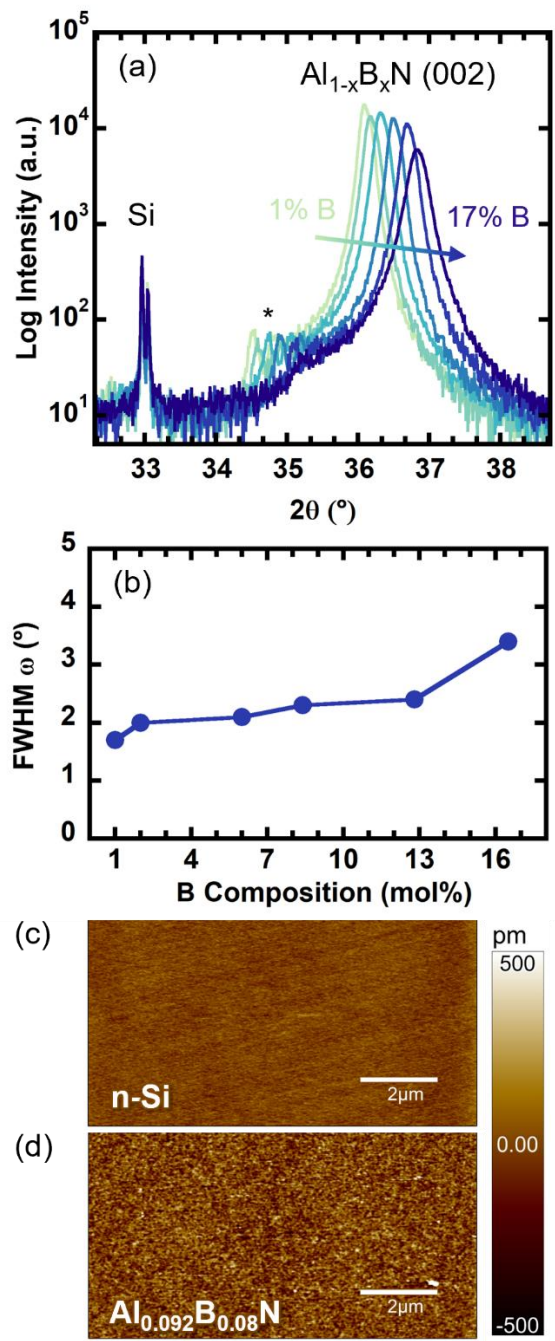

*Fig. 1.* (a) offset XRD θ-2θ scans of the 200 nm (002) $Al_{1-x}B_xN$ peak where x = <0.01, 0.02, 0.06, 0.08, 0.13, and 0.17, (b) rocking curve FWHM of the 2θ (002) peak as a function of B concentration for each sample, (c) AFM of as received n-type Si wafer surface roughness, and (d) 200 nm $Al_{0.92}B_{0.08}N$ surface roughness. Starred peaks in Figure 1a represent x-ray spectral lines.

Fig. 2 shows polarization hysteresis measurements collected from the B composition series using a 10 Hz triangle wave, bias is applied to the *n*-type Si substrate, and the W electrode is ground.

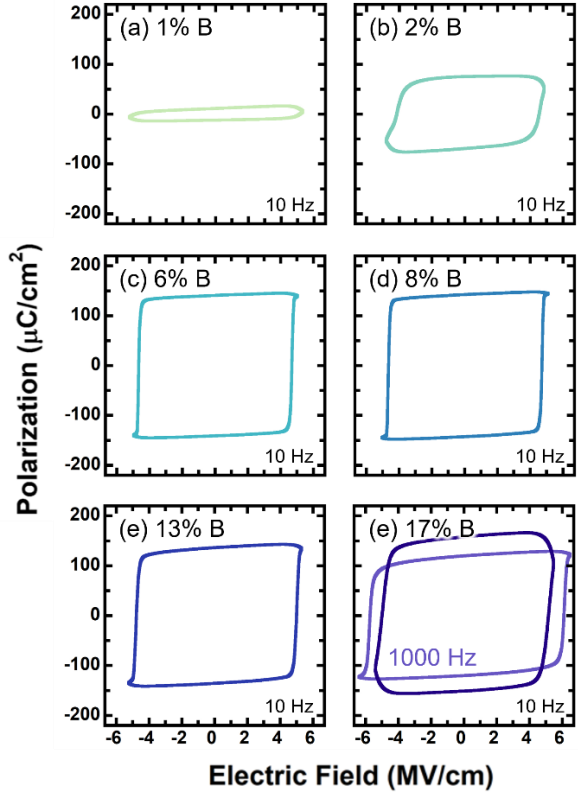

*Fig. 2. Polarization hysteresis progression of 200 nm $Al_{1-x}B_xN$ increasing B composition (a-f) of x ~0.01, 0.02, 0.06, 0.08, 0.13, and 0.17 mol %.*

Hysteresis loops with B content between <0.01 and 0.02 do not reach the expected remanent polarization range above 100 µC/cm². Additional electric fields only produce runaway leakage currents prior to breakdown. Figs. 2(c-e) shows fully saturated and high resistance loops for 6%, 8%, and 13% B. Remanent polarization values between 130-140 µC/cm² and coercive fields as low as 4.6 MV/cm are observed at 10 Hz. At 17% B (Fig. 2(f)) the material still switches, however, the loop becomes less square and leakage current contributions to the apparent polarization are evident. We choose 10 Hz for this measurement set so that leakage contributions are pronounced. For context, Fig. 2(f) also includes a loop for the same film collected at 1000 Hz, at this frequency the coercive field increases substantially – as anticipated – but despite the higher applied switching fields, the hysteresis loop shows more reasonable remanent polarization values and saturation at positive and negative electric fields. While it is well known that coercive fields are frequency dependent in ferroelectric crystals, the specific dependency has not been reported for the $Al_{1-x}B_xN$ system on Si. Fig. 3 shows hysteresis loops for available frequencies on an 8% B $Al_{1-x}B_xN$ capacitor showing four decades of dynamic range. From these room temperature measurements, we find a logarithmic relation between coercive field and frequency given by $E_c = 3.986 + 0.661 \log(f)$ [MV/cm].

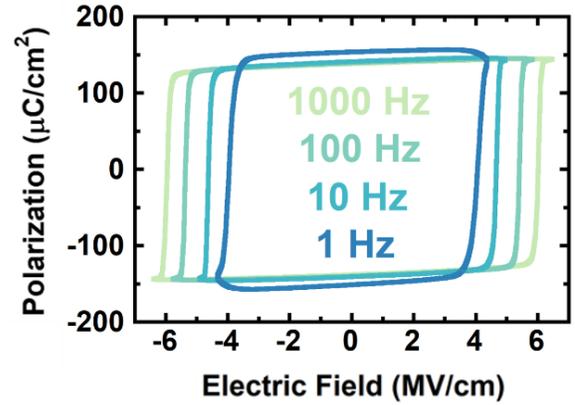

*Fig. 3. Frequency dependent hysteresis for an $Al_{0.92}B_{0.08}N$ film with 8% B substitution.*

As reported by Hayden et al. Si exposure time to the N2/Ar plasma influences the $Al_{1-x}B_xN$ crystallinity, texture, and electrical properties. To demonstrate this effect, three samples were prepared with varied plasma treatment time from 0, 15, and 30 minutes. Fig. 4(a) and Fig. 4(b) summarize the key findings: (1) longer plasma times produce lower crystalline mosaicity; (2) better crystallinity; (3) no difference in surface roughness; and (4) better insulation resistance and polarization hysteresis.

Consistent with the findings of Hayden et al., a longer exposure and thus more complete native oxide to nitride conversion facilitates both AlN growth and texture. It is also worth noting that all samples receiving a plasma treatment in this study have RMS roughness values at or below 400 pm.[5]

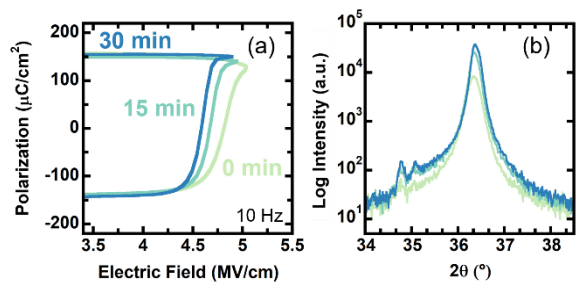

*Fig 4.* (a) Polarization hysteresis of 200 nm $Al_{0.94}B_{0.06}N$ grown after the n-type Si had no plasma treatment, 15 min plasma treatment, and 30 min plasma treatment, and (b) θ-2θ scans of the 200 nm (002) $Al_{1-x}B_xN$ intensity increasing with longer plasma treatment.

Microstructure analysis of an $Al_{0.92}B_{0.08}N$ film on a 30 min plasma treated *n*-type Si wafer by TEM is shown in Fig. 5(a) for the entire cross-section and in Fig. 5(b) at a region close to the bottom interface. In general, there is substantial diffraction contrast in all sections indicating a large concentration of defects and grain boundaries. In contrast to previous work growing $Al_{1-x}B_xN$ on W bottom electrodes where the defect density falls and the grain diameters increase as a function of distance away from the bottom interface, the microstructure in the present case is mostly uniform except for locations very close to the *n*-type Si substrate as shown in Fig. 5(b). It is interesting to note that despite the additional disorder in the present films relative to films grown on W, the high-field electrical properties are nearly identical, suggesting that such defects do not have a prominent influence.[5]

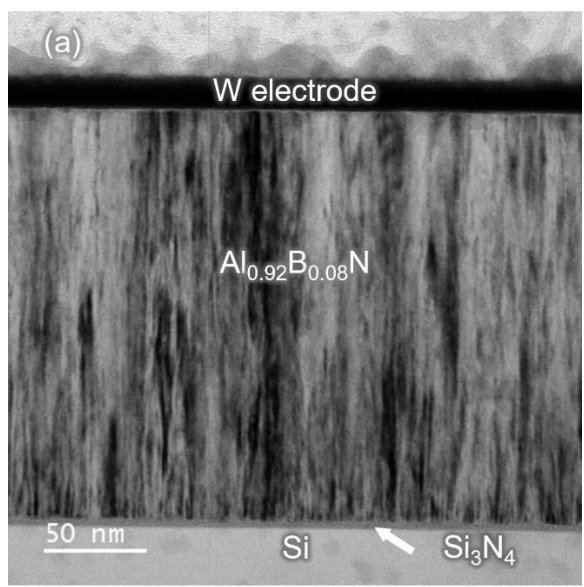

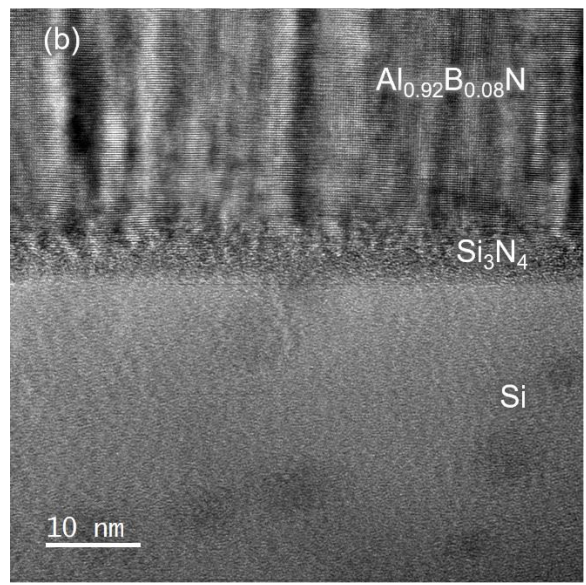

*Fig. 5.* (a) Cross-sectional bright field TEM of $Al_{0.92}B_{0.08}N$ on n-type Si, with a $Si_3N_4$ intermediate layer and (b) close up on $Si/Al_{1-x}B_xN$ interface.

EELS analysis is used to investigate the $Si/Al_{1-x}B_xN$ interfacial chemistry, as presented in Fig. 6. The results indicate that this interlayer is most likely $Si_3N_4$ due to $Si-L_{2,3}$ shifting to higher energy loss and oxygen content being below detection limits, whereas clear N signals are present along with Si. Ellipsometry analysis of this layer yields a refractive index of ~2.05 (at 633 nm) supporting that the interlayer is largely $Si_3N_4$. Current vs field (IE) measurement data in Fig. 6(d) of only plasma treated Si tested through the $Si_3N_4$ thickness is shown to be electrically insulating. In Hayden et al.'s previous plasma treatment analysis, XPS depth profiles showed a small amount of oxygen was present in the plasma treated interlayer.[5] Currently, it is not clear if full interface nitridation is important for electrical properties and crystallographic texture.

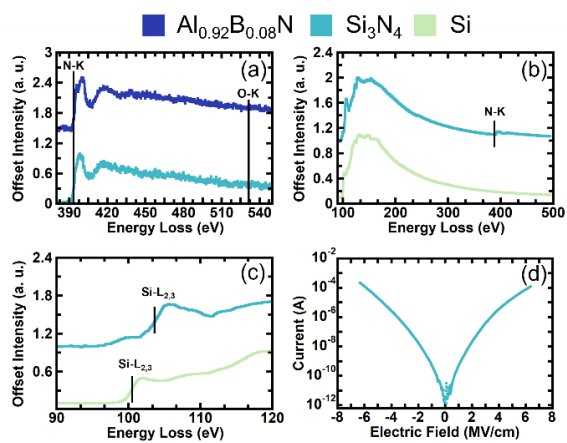

*Fig. 6.* EELS of 200 nm $Al_{0.92}B_{0.08}N$, $Si_3N_4$, and n-type Si showing (a) $Al_{0.92}B_{0.08}N$ and $Si_3N_4$ nitrogen K edge and no oxygen K edge, (b) Si and $Si_3N_4$ $L_{2,3}$ and nitrogen K edge (c) Si and $Si_3N_4$ $L_{2,3}$ edge, and (d) IV curve of 30 minute plasma treated $Si_3N_4$ on n-Si without an $Al_{1-x}B_xN$ layer.

Thickness scaling without changing deposition parameters is investigated not only to understand better the initial interface contributions but also to move closer to realistic device thicknesses. Fig. 7 shows polarization hysteresis and capacitance data for $Al_{1-x}B_xN$ films thickness scaled from 200 nm to 100 nm, 50 nm, and 25 nm. Leakage contributions to polarization begin to increase at 50 nm. For this experiment, 8% B samples are chosen because they show the overall lowest leakage at 200 nm layers. While films at 25 nm films are ferroelectric and switch by CV measurements at 20 kHz the leakage contributions are likely too large to support endurance properties that are device-compelling. Additional work is needed to understand the origins of this leakage trend but note that none of these films presented were capped with metal in vacuum. This is stressed due to the importance of capping scaled $Al_{1-x}Sc_xN$ films to prevent surface oxidation with the atmosphere.[12]

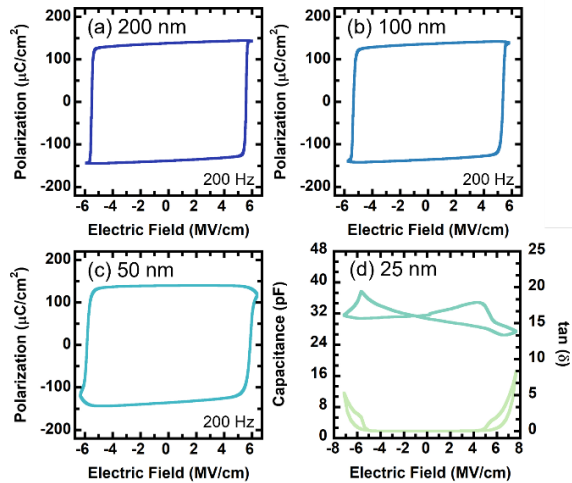

*Fig. 7.* Polarization hysteresis varying thickness of $Al_{0.92}B_{0.08}N$ on n-type Si at (a) 200 nm, (b) 100 nm, and (c) 50 nm, and (d) capacitance vs field butterfly-loop of a 25 nm film.

An interesting observation is that this thickness dependence is nearly identical to that previously observed for ferroelectric AlN by multiple authors on substrates that afford better crystallinity and texture.[5,15,16] This points to the possibility that leakage currents – especially at low thickness – may not be tied intimately to structural defects.[17,18]

## Conclusion

We have shown that well-oriented $Al_{1-x}B_xN$ thin films over a range of compositions can be grown directly on *n*-type Si, which serves as the bottom electrode. $Al_{1-x}B_xN$ grain size differs from previous reports while electrical properties remain consistent. With the aid of a plasma treatment before growth, a $Si_3N_4$ template that promotes $Al_{1-x}B_xN$ nucleation substantially enhances electrical quality. Growth has been achieved while maintaining back-end processing temperatures and by utilizing the substrate as the bottom electrode, hence, reducing processing steps and interfacial complexity. While mosaicity slightly increased compared to utilizing an AlN/W template bottom electrode, electrical properties are comparable to previous results even below 200 nm and at low frequency. Thickness scaling continues to become an important long-term goal, which elucidates interfacial and dislocation contributions, especially at the first 5 nm of the film. This work ultimately aims to accelerate the integration of $Al_{1-x}B_xN$ films onto commercially prevalent *n*-type Si with prospects of robust non-volatile memory and more streamlined manufacturing science.

## Acknowledgment

This work was supported by the center for 3D Ferroelectric Microelectronics (3DFeM[2]), an Energy Frontier Research Center funded by the U.S. Department of Energy, Office of Science, Office of Basic Energy Sciences Energy Frontier Research Centers program under Award No. DE-SC0021118.

## Author Declarations

### Conflicts of Interest

The authors have no conflicts to disclose.

### Author Contribution

**Ian Mercer:** Conceptualization (equal); Formal analysis (equal); Data curation (lead);); Investigation (lead); Methodology (equal); Validation (equal); Visualization (lead); Writing – original draft (lead); Writing – review & editing (lead).
**Chloe Skidmore:** Conceptualization (equal); Formal analysis (equal); Investigation (supporting); Methodology (equal); Validation (lead); Visualiza-tion (supporting); Writing – original draft (supporting); Writing – review & editing (supporting). **Jon-Paul Maria:** Conceptualization (equal); Formal analysis (supporting); Funding acquisition (lead); Methodology (equal); Project administration (lead); Resources (lead); Supervision (lead); Writing – original draft (supporting); Writing – review & editing (supporting). **Sebastian Calderon:** Conceptualization (supporting); Validation (supporting); Data Curation (supporting). **Elizabeth Dickey:** Funding acquisition (supporting); Methodology (supporting); Project administration (supporting); Resources (supporting); Writing – review & editing (supporting).

## Data Availability

The data that support the findings of this study are available from the corresponding author upon request.